\documentclass[final,authoryear,3p,times,twocolumn]{elsarticle}
\usepackage{graphicx}
\usepackage{fancyhdr}
\usepackage{eso-pic}

\newcommand\BackgroundPic{
\put(0,0){
\parbox[b][\paperheight]{\paperwidth}{%
\vfill
\centering
\vspace*{-12.7cm}
\vspace*{-12.55cm}
\hspace*{-0.3cm}
\includegraphics[width=150mm, angle=0, keepaspectratio]{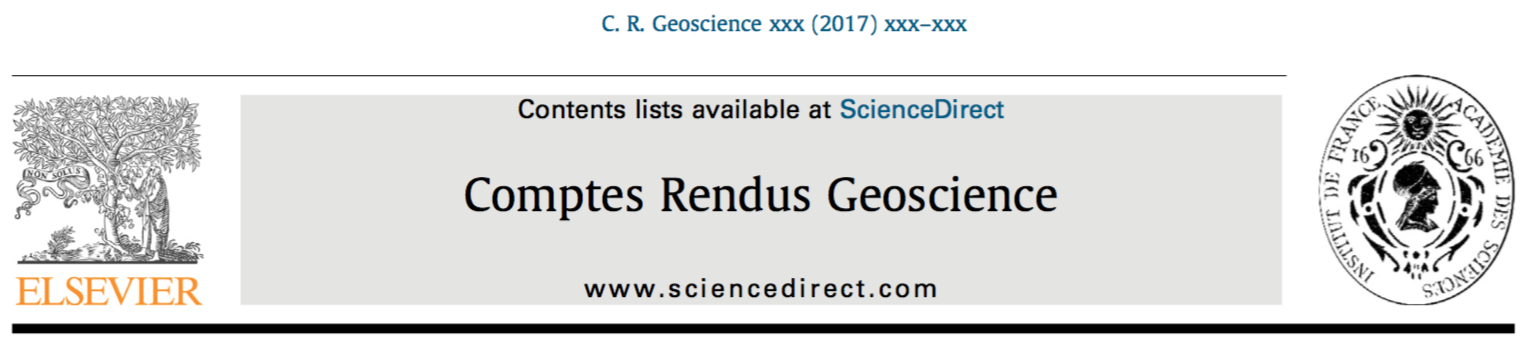}%
\vfill
}}}

\newcommand\BottomPic{
\put(0,0){
\parbox[b][\paperheight]{\paperwidth}{%
\vfill
\centering
\vspace*{24.5cm}
\includegraphics[width=168.5mm, angle=0, keepaspectratio]{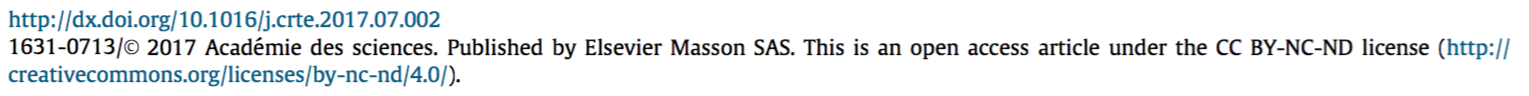}%
\vfill
}}}

\def\ltsima{$\; \buildrel < \over \sim \;$}
\def\simlt{\lower.5ex\hbox{\ltsima}}
\def\gtsima{$\; \buildrel > \over \sim \;$}
\def\simgt{\lower.5ex\hbox{\gtsima}}
\def\arcmin{$^\prime$}
\def\arcsec{$^{\prime\prime}$}

\newcommand*\aap{A\&A}

\newcommand*\ao{Appl.~Opt.}

\newcommand*\apj{ApJ}
\newcommand*\apjl{ApJ}

\newcommand*\apjs{ApJS}

\newcommand*\araa{ARA\&A}

\newcommand*\mnras{MNRAS}

\newcommand*\nat{Nature}

\usepackage{amssymb}

\journal{Comptes Rendus Geoscience}

\begin{document}

\AddToShipoutPicture*{\BackgroundPic}
\AddToShipoutPicture*{\BottomPic}

\begin{frontmatter}



\title{Interstellar filaments and star formation}


\author{Philippe Andr\'e}

\address{D\'epartement d'Astrophysique (AIM), 
CEA Saclay, 91191 Gif-sur-Yvette, France -- E-mail: pandre@cea.fr}

\begin{abstract}

Recent studies of the nearest star-forming clouds of the Galaxy at submillimeter wavelengths with the $Herschel$ Space Observatory have provided us with unprecedented images of the initial conditions and early phases of the star formation process. 
The $Herschel$ images reveal an intricate network of filamentary structure in every interstellar cloud. These filaments all exhibit remarkably similar widths - about a tenth of a parsec - but only the densest ones contain prestellar cores, the seeds of future stars. The $Herschel$ results favor a scenario in which interstellar filaments and prestellar cores represent two key steps in the star formation process: first turbulence stirs up the gas, giving rise to a universal web-like structure in the interstellar medium, then gravity takes over and controls the further fragmentation of filaments into prestellar cores and ultimately protostars. This scenario provides new insight into the origin of stellar masses and the star formation efficiency in the dense molecular gas of galaxies.  
Despite an apparent complexity, global star formation may be governed by relatively simple universal laws from filament to galactic scales. 

\end{abstract}


\end{frontmatter}


\section{Introduction}
\label{intro}

Star formation is one of the most complex processes in astrophysics, involving a subtle interplay between gravity, turbulence, 
magnetic fields, feedback mechanisms, heating and cooling effects \citep[e.g.][]{McKeeOstriker2007} etc...
Yet, despite this apparent complexity, the net products of the star formation process on global scales are relatively simple 
and robust. In particular, the distribution of stellar masses at birth or stellar initial mass function (IMF) is known to be 
quasi-universal  \citep[e.g.][]{Bastian+2010}.
Likewise, the star formation rate on both interstellar-cloud and galaxy-wide scales is related to the mass of dense molecular gas 
available by rather well defined ``star formation laws''  \citep[e.g.][]{Lada+2012, Shimajiri+2017}.

This paper presents an overview of recent observational results obtained with the {\it Herschel} Space Observatory 
\citep[][]{Pilbratt+2010} and other facilities on the texture of nearby star-forming clouds, which suggest that it may be possible to explain, 
at least partly, the IMF and the global star formation efficiency in the Galaxy in terms of the quasi-universal filamentary structure of the 
cold interstellar medium (ISM) out of which stars form.
Interestingly, 
the filamentary web of cold interstellar gas responsible for star formation within galaxies 
bears a marked resemblance to the cosmic web 
of intergalactic gas and dark matter leading to galaxy formation in cosmological simulations \citep[cf.][and 
\S ~4 below]{Springel+2005,Freundlich+2014}.

\begin{figure}[!h]
\center
\resizebox{8cm}{!}{     
\includegraphics[angle=0,scale=0.5]{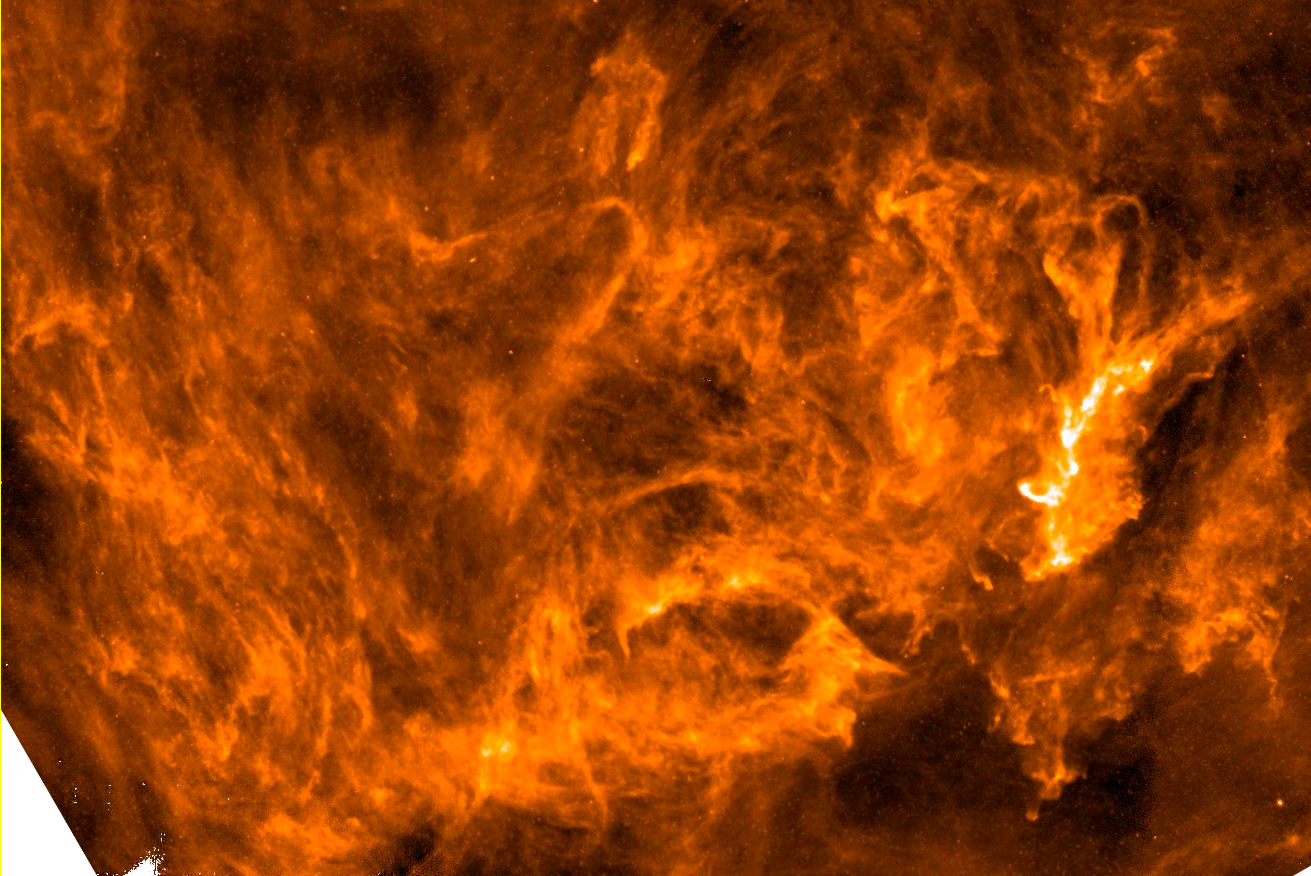}}
\caption{$Herschel$/SPIRE 250~$\mu$m dust continuum image of a portion of the Polaris flare translucent cloud 
($d \sim 150$~pc) taken as part of the HGBS survey \citep[e.g.][]{mamd+2010,Ward+2010,Andre+2010}.
Note the widespread filamentary structure. 
}
\label{polaris_filaments}
\end{figure}

\section{Summary of {\sc Herschel} results supporting a filamentary paradigm for star formation}

{\it Herschel} imaging surveys of Galactic star-forming regions 
have confirmed 
the ubiquitousness of filaments in Galactic molecular  clouds (see Fig.~\ref{polaris_filaments})  
and suggested an intimate connection between the filamentary structure 
of the ISM 
and the star formation process  \citep[e.g.][]{Andre+2010, Molinari+2010}. 
While molecular clouds have been known to exhibit large-scale filamentary 
structures for quite some time 
\citep[e.g.][and references therein]{Schneider+1979, Johnstone+1999, Myers2009}, 
{\it Herschel} observations now demonstrate 
that these filaments are truly ubiquitous in the cold ISM \citep[e.g.][]{Menshchikov+2010,Hill+2011,Wang+2015}, 
probably make up a dominant fraction of the dense gas in molecular clouds \citep[e.g.][]{Schisano+2014,Konyves+2015}, 
present a high degree of universality in their properties \citep[e.g.][]{Arzoumanian+2011}, 
and are the preferred birthplaces of prestellar cores \citep[e.g.][]{Konyves+2015,Marsh+2016}.
This means that interstellar filaments probably play a central role in the star formation process \citep[e.g.][]{Andre+2014}. 

 \begin{figure}[!h]
 \center
  \resizebox{8.0cm}{!}{         
\includegraphics[angle=0]{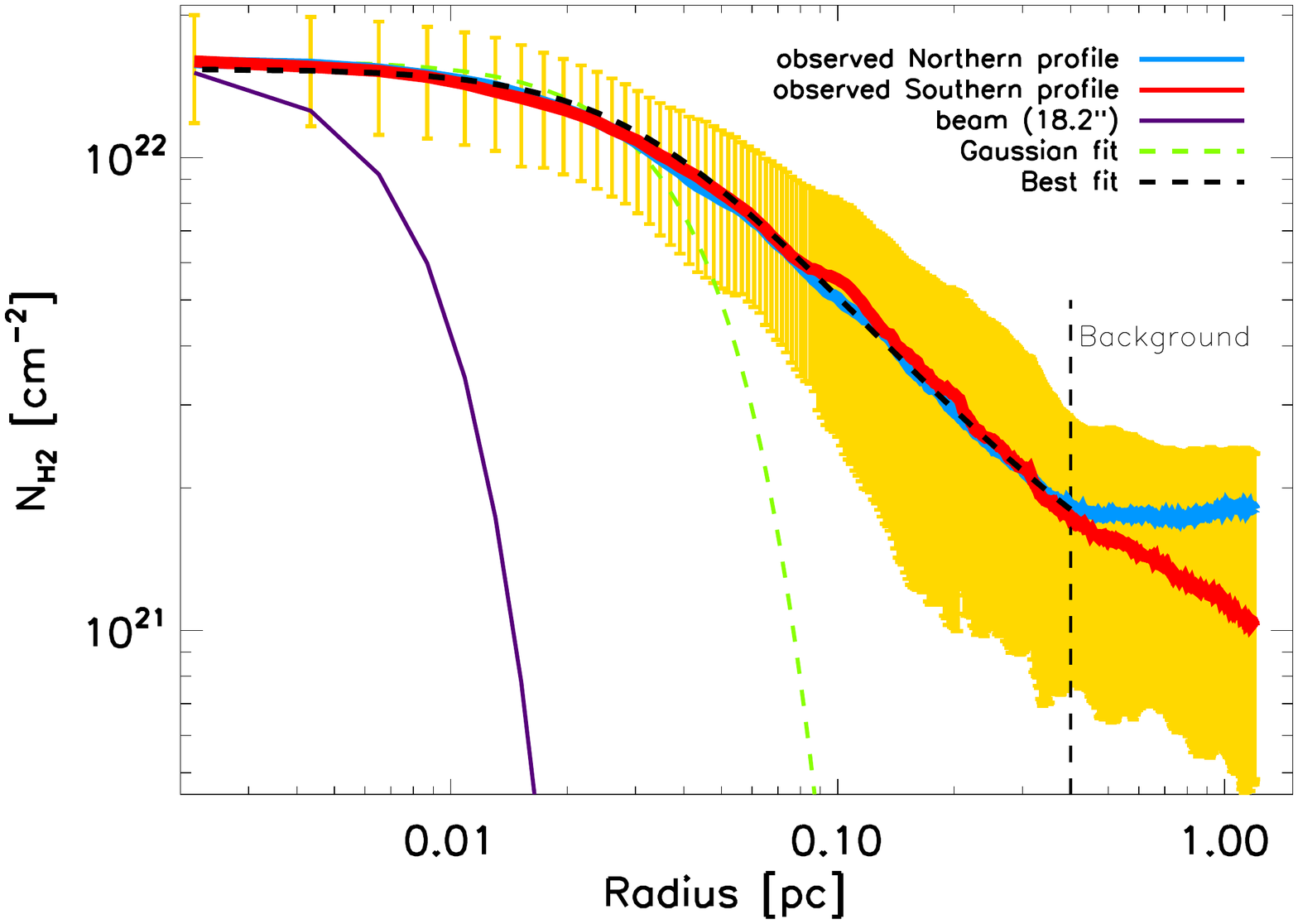}}  
   \caption{Mean radial column density profile observed with $Herschel$ 
perpendicular to the B213/B211 filament in Taurus, 
for both the Northern (blue curve) and the Southern part (red curve) of the filament.
The yellow area shows the ($\pm 1\sigma$) dispersion of the distribution of radial profiles along the filament.
The inner solid purple curve shows the effective 18\arcsec ~HPBW resolution
(0.012~pc at 140~pc) of  the column density map used to construct the profile. 
The dashed black curve shows the best-fit Plummer-like model 
of the form $N_p(r) = N_{\rm H2}^0/[1+ (r/R_{\rm flat})^2]^{\frac{p-1}{2}} $,
which has a power-law index $p$=2.0$\pm$0.4 and a diameter $2\times R_{\rm \rm flat} = 0.07 \pm 0.01$~pc.
It matches the data very well for $r$$\leq$0.4\,pc. 
(From Palmeirim et al. 2013.)
 }
              \label{fil_prof}
    \end{figure}
    
\begin{figure*}[!htp]
      \center
  \resizebox{13cm}{!}{         
\includegraphics[angle=0]{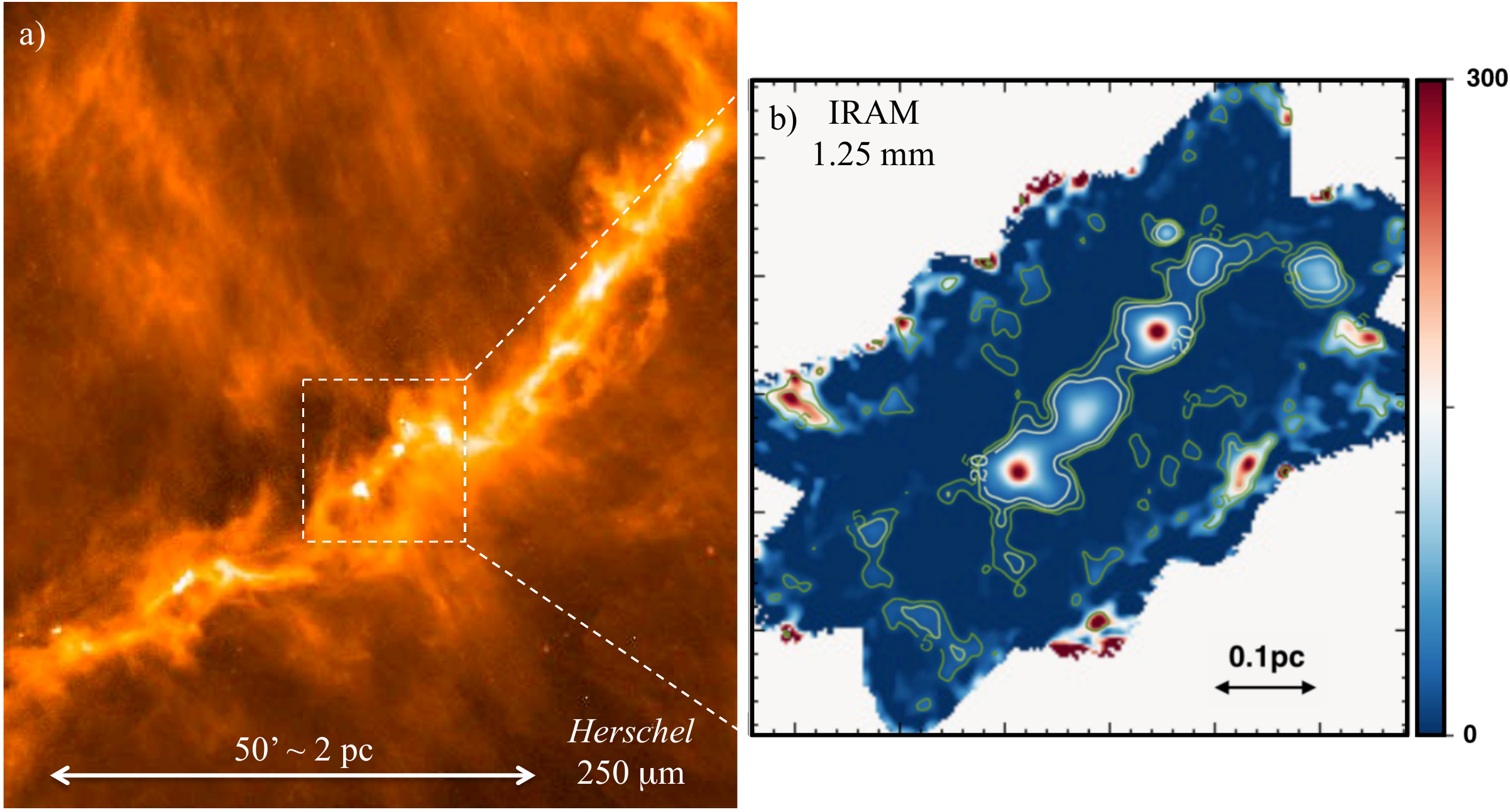}}  
   \caption{{\it (a)} {\it Herschel}/SPIRE 250~$\mu$m dust continuum image of the Taurus B211/B213 filament in the Taurus molecular cloud 
(Palmeirim et al. 2013, Marsh et al. 2016). Note the presence of several prominent dense cores along the filament. 
{\it (b)} IRAM/NIKA 1.25~mm dust continuum image of the central part of the {\it Herschel} field on the left (from \citealp{Bracco+2017}), 
showing at least three prominent dense cores.
 }
              \label{taurus_twins}
\end{figure*}    
   
\subsection{A characteristic filament width} 

Detailed analysis of the radial column density profiles derived from {\it Herschel} submm dust continuum data 
shows that, at least in the nearby clouds of the Gould Belt, 
filaments are characterized by a very narrow distribution of 
inner widths $W$ with a typical FWHM 
value $\sim 0.1\,$pc (much higher than the $\sim 0.01\,$pc resolution provided by {\it Herschel} at  the distance $\sim 140\,$pc 
of the nearest clouds)  and a dispersion lower than a factor of 2 
(\citealp{Arzoumanian+2011}; \citealp{Palmeirim+2013}; \citealp{KochRosolowsky2015}). 
Independent measurements of filament widths have generally been consistent with this {\it Herschel} finding 
when performed in submm continuum emission. For instance, \citet{Salji+2015} found an averaged 
deconvolved FWHM width of $0.08^{+0.07}_{-0.03}\,$pc for 28 filaments detected at $850\, \mu$m with SCUBA-2. 
It should be pointed out, however, that significant variations around the mean inner width of $\sim 0.1\,$pc 
(by up to a factor of $\sim 2$ on either side) sometimes exist along the main axis of a given filament 
\citep[e.g.][]{Juvela+2012,Ysard+2013}. 
We also note that the 
report  by \citet{Henshaw+2017} of a typical width of $\sim 0.03\,$pc for filamentary structures observed 
in the 1.1~mm dust continuum with ALMA toward the infrared dark cloud G035.39-00.33 may be 
affected by interferometric filtering of large-scale emission. 

Recently, \citet{Panopoulou+2017} raised the issue of whether the existence a characteristic filament width 
is consistent with the scale-free nature of the power spectrum of interstellar cloud images (well described by a single power law 
from $\sim 0.01\,$pc to $\sim 50\,$pc -- \citealp{mamd+2010,mamd+2016}). 
They suggested that the filament widths obtained by 
Gaussian fitting may be strongly correlated with the range of radii over which the filament profiles are fitted 
and may not indicate a genuine characteristic width. 
This suggestion is questionable for the following reasons. 
First, simulations indicate that the power spectra of synthetic cloud images including populations 
of simple model filaments, all 0.1~pc in diameter, remain 
consistent with the observed, 'scale-free' power spectra 
for realistic distributions of filament contrasts over the background (A. Roy et al., in prep.). 
Second, in the case of dense filaments whose radial profiles often feature power-law wings,  
a typical inner diameter of $\sim 0.1\,$pc is also found using a Plummer-like model function of the form 
$N_p(r) = N_{\rm H2}^0/[1+ (r/R_{\rm flat})^2]^{\frac{p-1}{2}} $, which provides a much better fit 
to the overall radial profiles than a simple Gaussian model \citep[see Fig.~\ref{fil_prof} adapted from][]{Palmeirim+2013}. 
Third, 
the range of radii used for the Gaussian fitting need not be 
arbitrarily fixed 
but can be 
adjusted depending on an initial estimate of the background level at each point along the crest of a filament \citep[cf.][]{Andre+2016}.
Finally, it is the {\it physical} inner width (in pc) that remains approximately constant from filament to filament 
in {\it Herschel} observations of nearby clouds, while the {\it angular} width (in arcsec) scales roughly as the inverse 
of the parent cloud's distance \citep{Arzoumanian2012}.
Although 
the width estimates reported for low-density filaments with a low contrast over 
the background are clearly more uncertain, we conclude that the median FWHM filament width of $0.09 \pm 0.04\, $pc 
reported by (\citealp{Arzoumanian+2011} -- see also Andr\'e et al., 2014) 
is most likely reflecting the presence of genuine 
characteristic scale corresponding to the inner diameter of filament systems, at least in the Gould Belt. 

The existence of this characteristic scale is challenging for numerical simulations of interstellar cloud turbulence 
\citep[e.g.][]{Smith+2014,Ntormousi+2016}, 
and the origin of the common inner width of interstellar filaments is currently debated. 
A possible interpretation is that filaments originate 
from planar intersecting shock waves due to supersonic interstellar turbulence \citep[e.g.][]{Pudritz+2013}
and that the filament width corresponds to the (magneto-)sonic scale below which the turbulence becomes subsonic in diffuse, 
non-star-forming molecular gas \citep[cf.][]{Padoan+2001,Federrath2016}. 
Alternatively, a characteristic width may arise if interstellar filaments are formed as quasi-equilibrium structures 
in pressure balance with a typical ambient ISM pressure $P_{\rm ext} {\sim} 2$$-$5$\times$$10^4 \, \rm{K\, cm}^{-3} $ 
(\citealp{Fischera+2012}; S. Inutsuka, private communication). 
A second alternative explanation relies on the thermodynamical properties of cold ISM gas \citep[e.g.][]{Larson2005,Hocuk+2016}
and connects the common inner width of {\it Herschel} filaments to the thermal Jeans length at the transition density 
$n_{H_2} \sim 10^4$--$10^5\, \rm{cm^{-3}}$ at which the effective equation of state of the gas $P \propto \rho^{\gamma_{\rm eff}}$
changes from $\gamma_{\rm eff} < 1$ (below the transition density) to $\gamma_{\rm eff} > 1$ (above the transition density). 
Yet another possibility is that the characteristic inner width of filaments may be set by the dissipation mechanism 
of magnetohydrodynamic (MHD) waves due to ion-neutral friction 
\citep[][]{Hennebelle2013,HennebelleAndre2013,Ntormousi+2016}. 
Recently, \citet{Basu2016} and \citet[][]{Auddy+2016} 
pointed out that the magnetized character 
of dense molecular filaments (see \S~6 below) implies that they cannot be true cylinders 
and that their 3D configuration should be closer to 
magnetic ribbons, i.e., triaxial objects flattened along the direction of the large-scale magnetic field.
On this basis, they developed an analytic model in which the lateral width of the ribbon-shaped filaments 
is independent of density, but their thickness corresponds to the Jeans scale, and a relatively flat 
distribution of apparent filament widths results from projection effects assuming a random set of viewing angles. 

%
\begin{figure*}[!htp]
      \center
  \resizebox{13cm}{!}{
\includegraphics[angle=0]{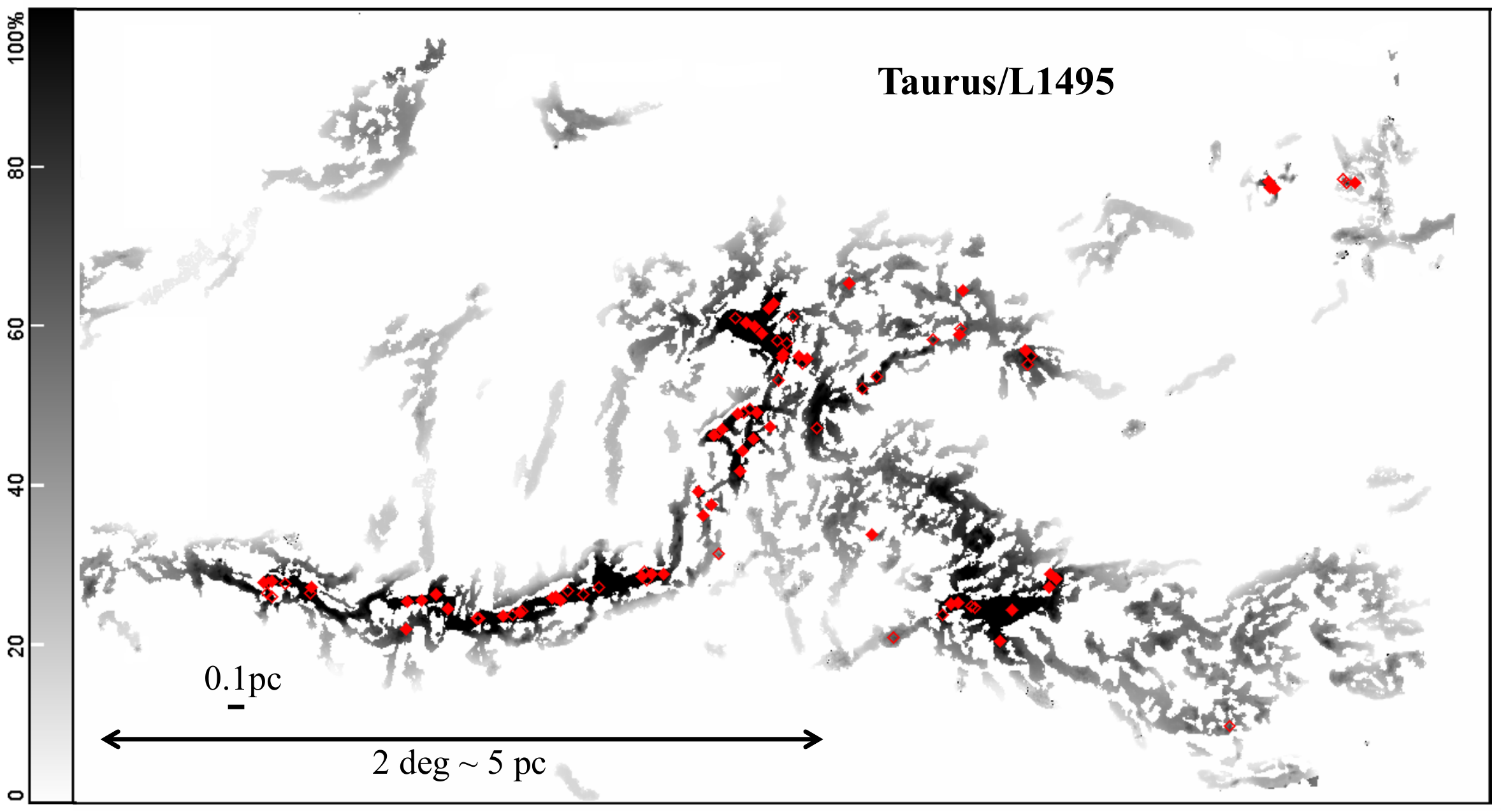}}    
\caption{Spatial distribution of prestellar cores (red diamonds) overlaid on a filtered 
column density map of the L1495 region in the Taurus molecular cloud, derived from 
{\it Herschel} Gould Belt survey data. To enhance the contrast of the filamentary structure, 
the original column density map was filtered using the multi-scale decomposition 
algorithm {\it getfilaments} \citep[][]{Menshchikov2013}.  
(Here, for better visualization of the filaments, only transverse spatial scales up 
to 0.1~pc were retained.)
Note how prestellar cores are only seen toward dense filamentary structures 
with mass per unit length in excess of $\sim 16\, M_\odot \, {\rm pc}^{-1} $.
(From Marsh et al. 2016.)
}
\label{taurus_cores}
\end{figure*}

\subsection{Link between dense cores and filaments} 

\begin{figure}[!h]
\center
\resizebox{8.0cm}{!}{     
\includegraphics[angle=0,scale=0.5]{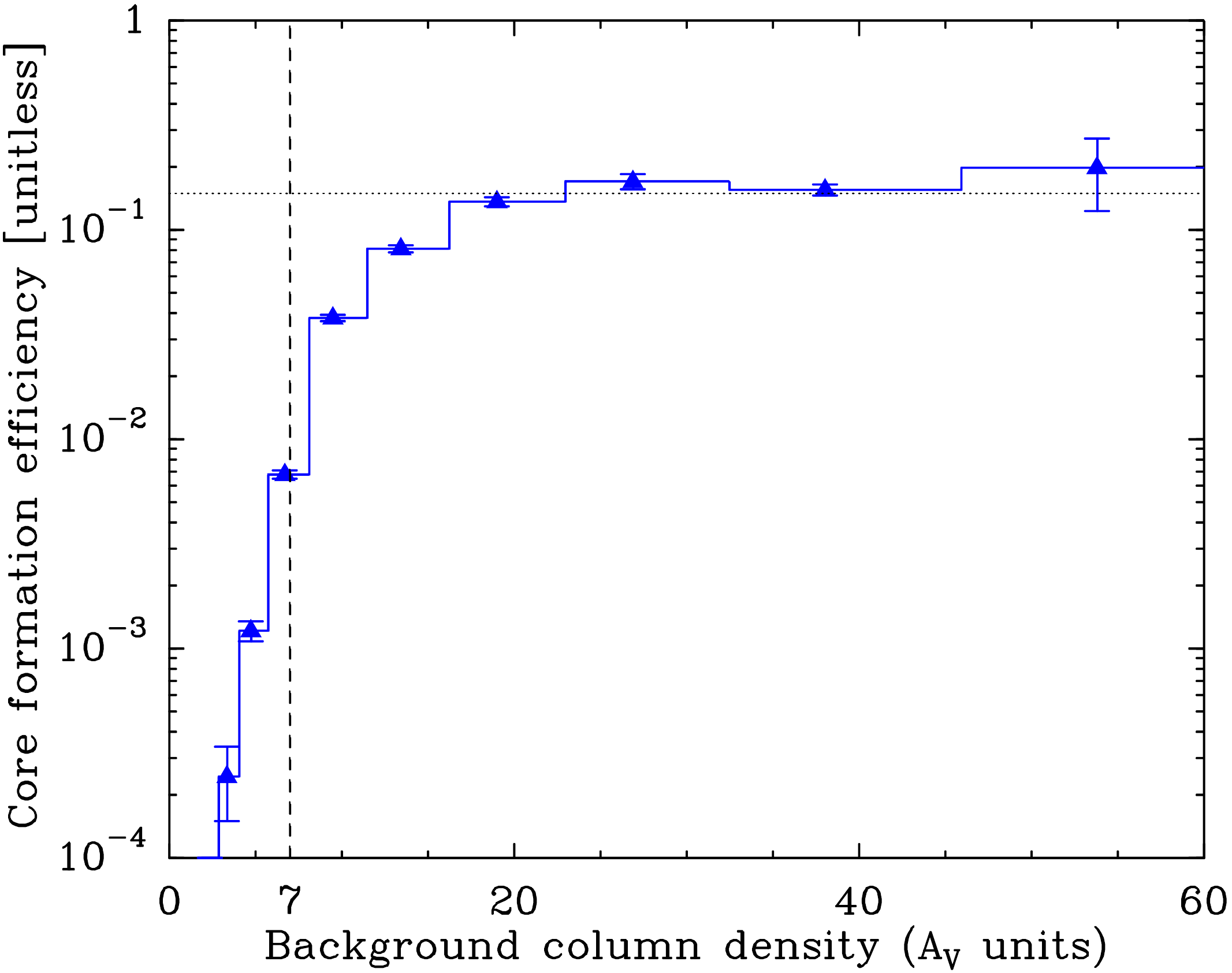}}
\caption{Plot of the differential core formation efficiency (CFE) 
in the Aquila complex 
as a function of background column density expressed in $A_{\rm V}$ units 
(blue histogram with error bars). The CFE was obtained by dividing the mass in the form of prestellar cores 
in a given column density bin by the cloud mass observed in the same column density bin. 
The vertical dashed line marks a fiducial threshold at   $A_V^{\rm back} \sim 7$. 
The horizontal dotted line marks the rough asymptotic value $\sim 15\% $  of the CFE  at $A_{\rm V} >  15$. 
(From K\"onyves et al. 2015.)
}
\label{aquila_diff_cfe}
\end{figure}

\begin{figure}[!h]
\center
\resizebox{7.5cm}{!}{     
\includegraphics[angle=0,scale=0.5]{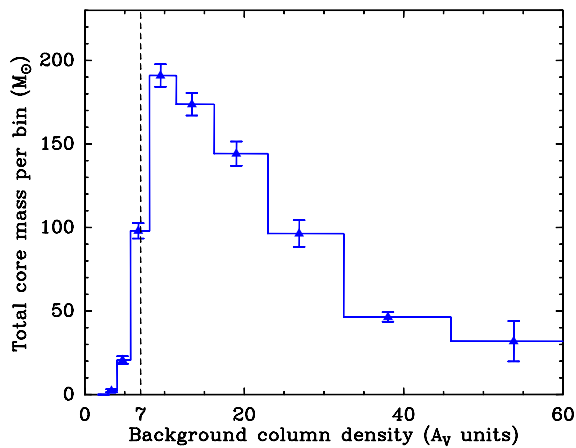}}
\caption{Mass in the form of prestellar cores in the Aquila molecular cloud 
as a function of background column density 
in $A_{\rm V}$ units.
The peak seen at $A_{\rm V} \sim 10$ implies that most prestellar cores form just above the column density threshold ($A_V^{\rm back} \sim 7$), 
in marginally supercritical filaments. 
(Adapted from K\"onyves et al. 2015.)
}
\label{aquila_diff_core_mass}
\end{figure}

Another major result from {\it Herschel} in nearby clouds is that most ($> 75\%$) 
prestellar cores 
are found to lie within dense filaments with column densities $N_{H_2}^{\rm fil} \simgt 7 \times 10^{21}\, {\rm cm}^{-2}$ 
\citep[e.g.][Marsh et al. 2016 -- see Figs.~\ref{taurus_twins} and~\ref{taurus_cores}]{Andre+2010,Konyves+2015}. 
Independent, albeit more indirect, evidence that prestellar cores are closely associated with filaments is provided by 
recent SCUBA-2 imaging observations which show that the spatial distribution of dense cores 
as traced by minimal spanning trees is itself filamentary \citep[e.g.][]{Kirk+2016,Lane+2016}, 
and highly correlated with the dense filaments detected with {\it Herschel} 
(K\"onyves et al., in prep.).  
Quantitatively, $\simgt 80\% $ of the candidate prestellar cores identified with {\it Herschel} lie within a radius of $< 0.1\, $pc 
of the crest of their nearest filament (\citealp{Konyves+2015}; \citealp{Bresnahan+2017}; Ladjelate et al., in prep.). 
The column density threshold above which prestellar cores are found in filaments is quite pronounced, 
as illustrated in Fig.~\ref{aquila_diff_cfe} which shows the observed core formation efficiency 
$ {\rm CFE_{obs}}(A_{\rm V}) = \Delta M_{\rm cores}(A_{\rm V})/\Delta M_{\rm cloud}(A_{\rm V}) $ 
as a function of the ``background'' column density of the parent filaments in the Aquila cloud complex. 
As can be seen, the plot of Fig.~\ref{aquila_diff_cfe} resembles a smooth step function.
The combination of this threshold with the typical column density probability function or 
distribution of cloud/filament mass as a function of column density 
\citep[][]{Kainulainen+2009,Lada+2010,Schneider+2013} 
implies that most prestellar cores form just above the column density threshold (cf. Fig.~\ref{aquila_diff_core_mass}).  

There is a natural interpretation of the column density threshold for prestellar core formation 
within filaments in terms of simple theoretical expectations for the gravitational instability of nearly isothermal gas cylinders. 
Given the typical width $W_{\rm fil} \sim 0.1$~pc measured for filaments  
and the relation $M_{\rm line} \approx \Sigma_0 \times W_{\rm fil}$ between the central 
gas surface density $\Sigma_0$ and the mass per unit length $M_{\rm line}$ of a filament, 
the threshold at $A_V^{\rm back} \sim 7$ or $\Sigma_{\rm gas}^{\rm back} \sim $~150~$M_\odot \, {\rm pc}^{-2} $ 
corresponds to within a factor of $<< 2$ to the critical mass per unit length 
$M_{\rm line, crit} = 2\, c_s^2/G \sim 16\, M_\odot \, {\rm pc}^{-1} $  
of isothermal long cylinders in hydrostatic equilibrium for a sound speed $c_{\rm s} \sim 0.2$~km/s, i.e., 
a typical gas temperature $T \sim 10$~K \citep[e.g.][]{Ostriker1964}. 
Therefore, as pointed out by \citet{Andre+2010}, 
the prestellar core formation threshold approximately corresponds to the line mass threshold 
above which interstellar filaments become gravitationally unstable to radial contraction and fragmentation along 
their length 
\citep[cf.][]{Inutsuka+1992,Inutsuka+1997}.

Some departures from the predictions of standard cylinder fragmentation theory are observed, however.
Linear fragmentation models for infinitely long, isothermal equilibrium  cylinders 
predict a characteristic core spacing of $4\, \times\, \,$the filament diameter or $\sim 0.4\,$pc \citep[e.g.][]{Inutsuka+1992}. 
In contrast, recent high-resolution 
studies \citep[e.g.][]{Takahashi+2013,Teixeira+2016,Kainulainen+2013,Kainulainen+2017} 
provide evidence of two-level hierarchical fragmentation within 
thermally supercritical filaments, with two fragmentation modes: 
a) a ``cylindrical'' fragmentation mode corresponding to groups of cores separated by $\sim 0.3\,$pc,
and b) a ``spherical'', Jeans-like fragmentation mode corresponding to a typical spacing $\simlt 0.1\,$pc 
between cores (and within groups). 
Moreover, the second mode appears to dominate since 
the median projected spacing observed between prestellar cores ($\sim 0.08\,$pc in Aquila -- K\"onyves et al. 2015) 
is consistent with the thermal Jeans length within critical filaments of $0.1\,$pc inner diameter at 10~K.

These findings have stimulated recent theoretical work to try and account for two fragmentation modes 
based on modifications to the standard cylinder fragmentation theory. 
\citet{Clarke+2016} explored the properties of non-equilibrium, accreting filaments and 
\citet{Gritschneder+2017} investigated the role of geometrical effects (i.e., the fact that real filaments 
are not perfectly straight but often exhibit some bends). 
While further work is still needed to get a perfect match with observations, 
it seems likely that a relatively simple model of filament fragmentation (including, for instance, the effects
of low levels of turbulence -- \citealp{Clarke+2017}) may be able to explain the observed distribution 
of core spacings along filaments.

Overall, the observational results summarized in this section 
support a filamentary paradigm for solar-type star formation,  
in two main steps \citep[cf.][]{Andre+2014}: 
First, large-scale compression of interstellar material in supersonic MHD flows 
generates a cobweb of filaments $\sim 0.1$ pc in width in the ISM; 
second, the densest filaments fragment into 
prestellar cores (and subsequently protostars) by gravitational instability above $M_{\rm line, crit} $. 
This paradigm differs from the classical ``gravo-turbulent'' fragmentation picture 
\citep[e.g.][]{Padoan+2002,MacLowKlessen2004,Hennebelle2008} 
in that it relies on the unique properties 
of filamentary geometry, such as the existence of a critical mass per unit length for nearly isothermal filaments. 

\section{Merits of the filamentary paradigm of star formation}

The filamentary paradigm of star formation sketched at the end of \S~2  
has several merits. 
One of them is that it provides a simple explanation for the origin of the lognormal ``base'' of the stellar initial mass 
function (IMF), for stellar masses around $\sim 0.3\, M_\odot $ \citep[cf.][]{Chabrier2005, Bastian+2010}.

\begin{figure}[!h]
\center
\resizebox{7.5cm}{!}{     
\includegraphics[angle=0,scale=0.5]{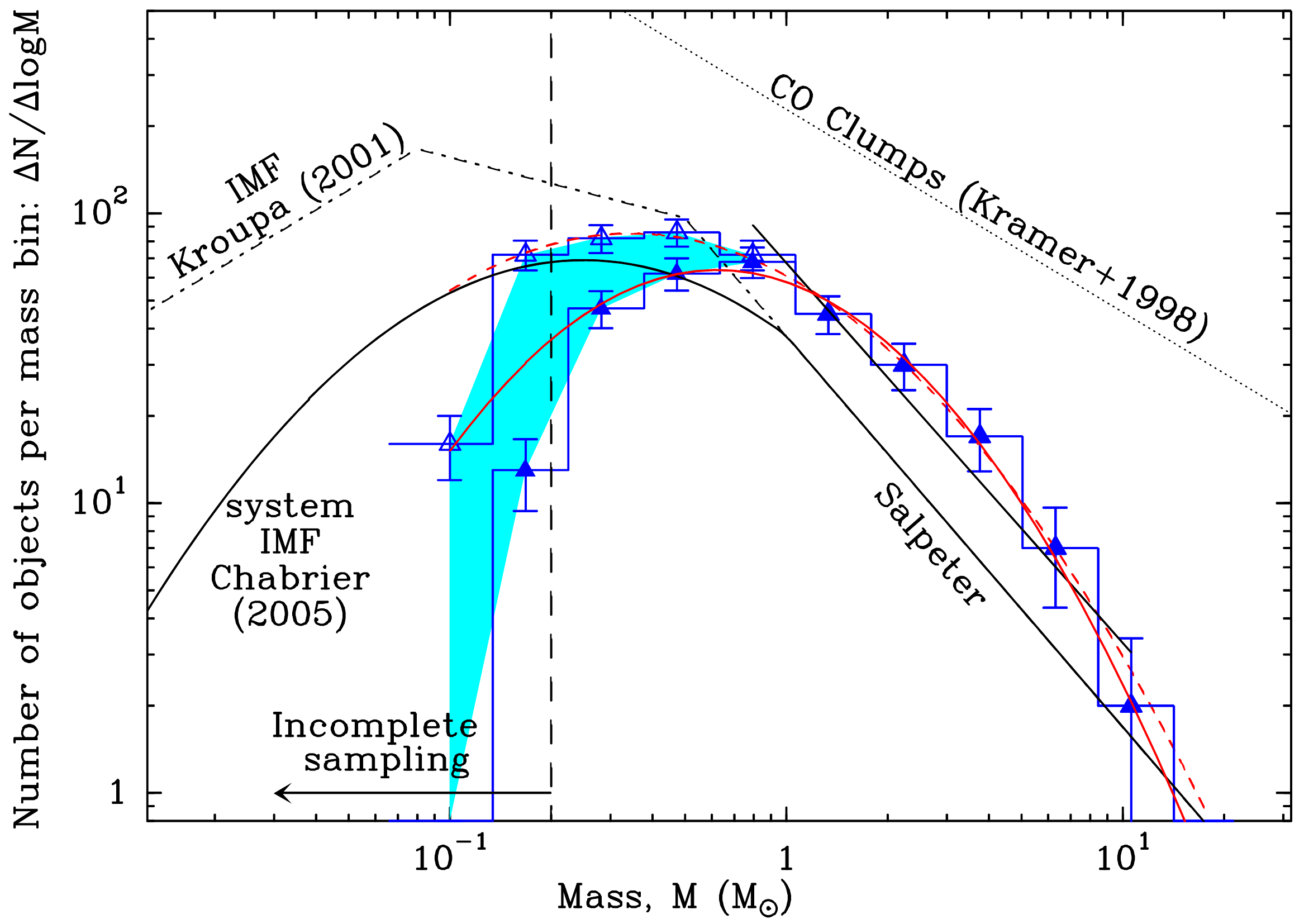}}
\caption{Observed core mass function (d$N$/dlog$M$) for 
the 446 candidate prestellar cores (upper blue histogram and triangles), and 292 robust prestellar cores (lower blue histogram and triangles) 
identified with {\it Herschel} in the Aquila cloud \citep[e.g.][]{Konyves+2015}. 
The shaded area in light blue reflects the uncertainties in the prestellar CMF. 
Lognormal fits (red curves)
and a power-law fit to the high-mass end of the CMF (black solid line) are superimposed. 
(Adapted from K\"onyves et al., 2015.)
}
\label{aquila_cmf}
\end{figure}

Indeed, as most prestellar cores appear to form in filaments just above the threshold for cylindrical gravitational instability, 
one expects that there should be a characteristic prestellar core mass corresponding to the local Jeans (or Bonnor-Ebert) mass
in marginally critical filaments. 
The thermal Bonnor-Ebert mass  
within $\sim 0.1$-pc-wide critical filaments at $\sim 10\,$K with $ M_{\rm line} \approx M_{\rm line, crit} \sim 16\, M_\odot \, {\rm pc}^{-1} $ 
and surface densities $\Sigma_{\rm cl}  \approx \Sigma_{\rm gas}^{\rm crit} \sim 160\, M_\odot \, {\rm pc}^{-2} $ 
is $ M_{\rm BE}  \sim 1.3\, c_s^4 /G^2 \Sigma_{\rm cl} \sim 0.5\, M_\odot $. 
This is very close to the peak of the prestellar core mass function (CMF) at $ \sim 0.6\, M_\odot $ as observed in the Aquila cloud 
complex (K\"onyves et al. 2015 -- cf. Fig.~\ref{aquila_cmf}). 
Moreover the observed prestellar CMF closely resembles the IMF, apart from a global shift in mass scale corresponding 
to an efficiency factor $\epsilon_{\rm core} \sim \,$30--50$\%$ \citep{Alves+2007,Konyves+2015}, 
which can be attributed to mass loss due to the effects of outflows during the protostellar phase 
\citep{Matzner+2000,Machida+2012}. 
The {\it Herschel} results therefore support the view that gravitational fragmentation of filaments is the dominant physical 
mechanism generating prestellar cores and stars around the peak of the CMF/IMF.

Naively, one would expect gravitational fragmentation to result in a narrow prestellar CMF sharply peaked at the median thermal
Jeans mass. However, a Salpeter-like power-law tail 
can quickly develop if turbulence has generated a field of initial density fluctuations within the filaments in the first place \citep{Inutsuka2001}. 
To investigate this possibility, \citet{Roy+2015}  carried out a statistical analysis of the line-mass fluctuations observed with {\it Herschel} along 
a sample of 80 subcritical or marginally supercritical filaments in three clouds of the Gould Belt. 
They found that the  
line-mass fluctuations are in the linear regime 
and that their power spectrum 
is well fit by a power law [$P(k) \propto k^\alpha$ with $\alpha = {-1.6\pm0.3}$], 
which is consistent with the 1D power spectrum generated by subsonic Kolmogorov turbulence  ($\alpha = -5/3$). 
Starting from such an initial power spectrum of line-mass fluctuations, the theoretical analysis by Inutsuka (2001) 
shows that the density perturbations quickly evolve -- in about two free-fall times or $\sim 0.5\,$Myr for a critical 0.1~pc-wide filament --
from a mass distribution similar to that of CO clumps \citep[cf.][]{Kramer+1998} 
to a population of protostellar cores whose mass distribution approaches the Salpeter power law at the high-mass end.
This occurs because
small-scale perturbations 
(i.e., small core masses) 
grow faster than large-scale perturbations, 
resulting in a steepening of the mass distribution with time.

\section{Formation of interstellar filaments: Observational constraints}

Broadly speaking, three classes of models have been proposed to explain the origin of interstellar filaments 
depending on whether gravity, supersonic turbulence, or magnetic fields play the primary role in generating 
the filamentary structure. 
When gravity dominates, e.g., in strongly Jeans-unstable clouds, it is known to amplify initial anisotropies, 
leading to large-scale collapse first along the shortest axis,  creating sheets or pancakes, then along the second dimension 
producing filaments, and finally 
ending up with 
spheroidal clumps/cores \citep[e.g.][]{Lin+1965,Zeldovich1970}. 
This is the favored mechanism to explain the growth of large-scale structure in the Universe and the formation 
of filaments in the cosmic web \citep{ShandarinZeldovich1989,Bond+1996}, 
and it has been proposed 
to operate in molecular clouds as well \citep[e.g.][]{Burkert+2004,Gomez+2014}. 
Observationally, there is little doubt that gravity is indeed the main player in strongly self-gravitating ``hub-filament'' systems, 
where a cluster-forming hub is observed at the center of a converging network of filaments (Myers 2009). 
Such systems are quite spectacular in high-mass star-forming regions 
(e.g. MonR2: \citealp{Didelon+2015,Pokhrel+2016}; SDC335: \citealp{Peretto+2013}), 
but also exist in clouds forming mostly (or only) low- to intermediate-mass stars 
(e.g. B59: \citealp{Peretto+2012}; L1688: Ladjelate et al., in prep.). 
Clearly, however, filaments are already widespread in gravitationally unbound clouds such as atomic (HI) clouds 
\citep[e.g.][]{McClure-Griffiths+2006,Clark+2014,Kalberla+2016} 
or translucent molecular clouds \citep[e.g. Polaris:][-- see Fig.~\ref{polaris_filaments}]{mamd+2010,Menshchikov+2010}, 
where the dominant role of gravity is difficult to invoke.
At least in these clouds, the filamentary structure must arise from other mechanisms, and the role of large-scale supersonic 
flows (turbulent or not) is likely. There is growing evidence of the presence of large-scale supersonic streams in the 
nearby ISM, possibly resulting from the action of stellar winds \citep[e.g.][]{BouyAlves2015}. 
Each of these large-scale
supersonic flows generates a locally planar shock wave which compresses interstellar matter. 
A filament is naturally produced at the intersection of two such planar shock waves \citep{Pudritz+2013},  
or as a result of the focusing effect generated by a single, but deformed MHD shock wave \citep{Inoue+2013}. 
Direct evidence of the role of large-scale compressive flows has been found with {\it Herschel} in the Pipe Nebula 
in the form of filaments with asymmetric column density profiles which most likely result from compression by the winds of the 
neighboring  
Sco OB2 association \citep{Peretto+2012}. 

\begin{figure}[!h]
\center
\resizebox{7.5cm}{!}{     
\includegraphics[angle=0,scale=0.5]{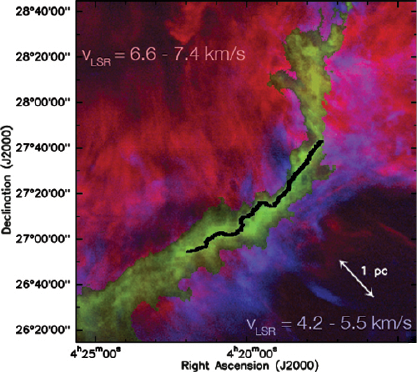}}
\caption{Transverse velocity gradient observed in $^{12}$CO(1--0) across the Taurus B211/B213 filament  in Taurus 
(Palmeirim et al. 2013, based on CO data from \citealp{Goldsmith+2008}). 
Note how the redshifted CO emission (in red) is mostly seen to the north-east and the blueshifted CO emission (in blue)
is mostly seen to the south-west of the filament (displayed in green).
}
\label{b211_vel_grad}
\end{figure}

Recent molecular line studies of the projected velocity field observed toward several nearby filament systems set interesting 
constraints on the origin of dense star-forming filaments. Both velocity gradients along and perpendicular to the major axis 
of molecular filaments have been detected \citep[e.g.][]{Peretto+2006,Kirk+2013,Palmeirim+2013}, 
but the gradients across filaments 
often dominate (e.g. \citealp{FernandezLopez+2014}; Mundy et al., in prep.). 
Such transverse velocity gradients, sometimes seen 
consistently along the entire length of the filamentary structure (see Fig.~\ref{b211_vel_grad}),  would not be observed 
if the filaments were embedded in (and accreting from) a spherical or cylindrical ambient cloud. 
They are most readily explained if the filaments are forming and growing inside sheet-like structures. 
Numerical MHD simulations of filament formation within shock-compressed layers generated by large-scale supersonic flows 
can reproduce this kinematic pattern \citep[][]{ChenOstriker2014,Inutsuka+2015}. 

\section{Evolution of molecular filaments: Observational constraints}

Locally at least, i.e., close to the filament axis, it appears that the transverse velocity gradients observed toward 
supercritical filaments are driven by gravity. Their magnitude is indeed roughly consistent with what is expected from 
accretion of background cloud material given the gravitational potential of the central filaments 
(e.g. Palmeirim et al. 2013; Mundy et al., in prep.; Shimajiri et al., in prep.). 
In the case of the Taurus B211/B213 filament, the estimated accretion rate of 27--50$\, M_\odot $/pc/Myr  
is such that it would take $\sim \, $1--2~Myr for the filament to double its mass (Palmeirim et al. 2013).
Another indication that thermally supercritical filaments are accreting background cloud material is the detection 
in dust and/or CO emission 
of low-density striations or sub-filaments perpendicular to the main filaments, and apparently feeding them from  the side. 
Examples include the Musca filament (Cox et al. 2016), the Taurus B211/B213 filament in Taurus (Goldsmith et al. 2008, Palmeirim et al. 2013), 
the Serpens-South filament in Aquila (e.g. Kirk et al. 2013), and the DR21 ridge in Cygnus~X \citep[][]{Schneider+2010,Hennemann+2012}.
The conclusion that supercritical filaments are generally accreting mass from their parent cloud is independently supported 
by the fact that their internal velocity dispersion is observed to increase with central column density, which is suggestive of an 
increase in virial mass per unit length with time \citep[][]{Arzoumanian+2013}. 

\begin{figure}[!h]
\center
\resizebox{7.5cm}{!}{     
\includegraphics[angle=0,scale=0.5]{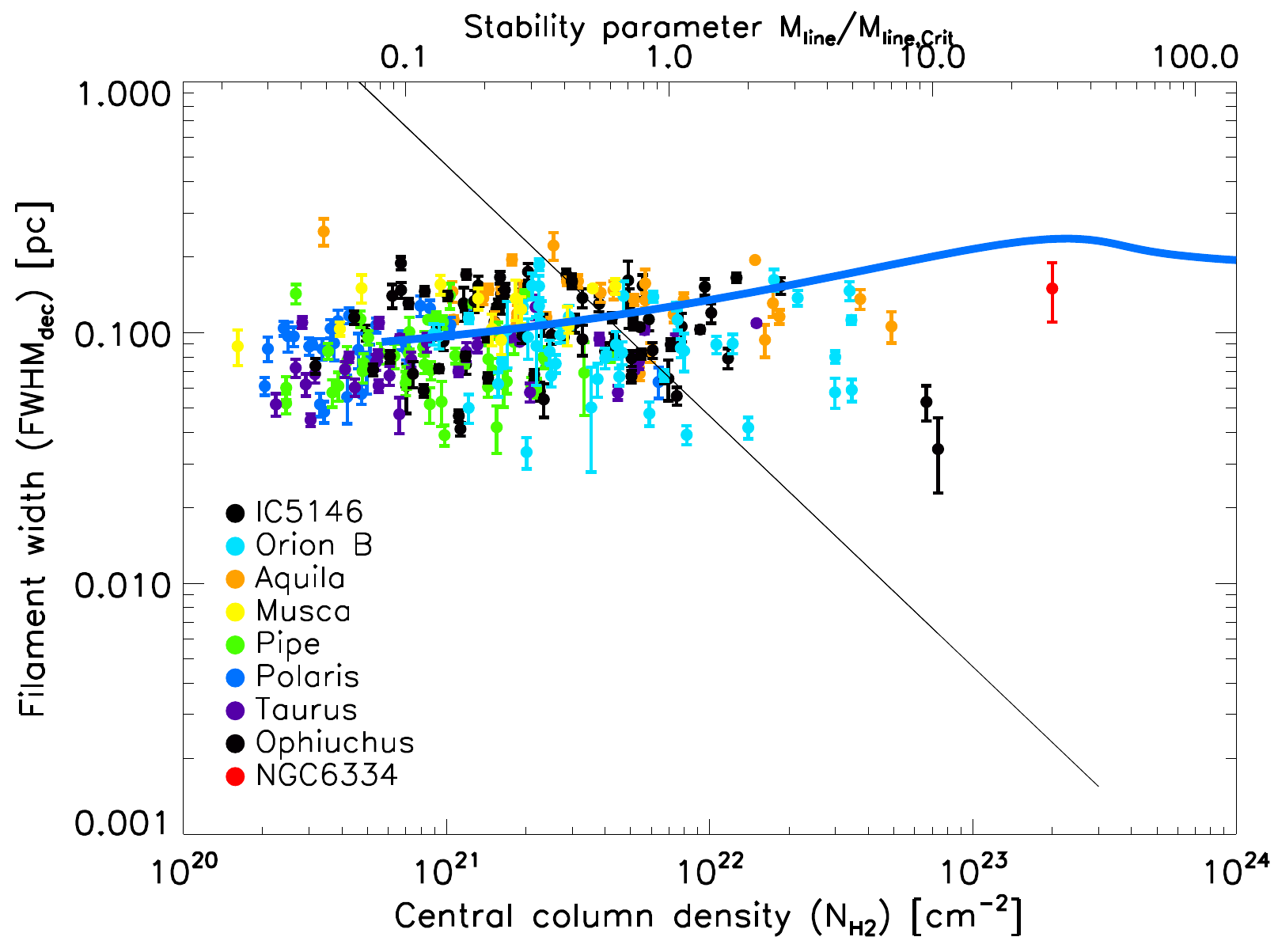}}
\caption{Plot of filament width against filament central column density in the accretion-driven MHD turbulence 
model proposed by \citet{HennebelleAndre2013} (blue curve). The data points with error bars correspond 
to width measurements for 278 nearby filaments based on {\it Herschel} observations \citep[][Arzoumanian et al., in prep]{Arzoumanian+2011}. 
The black straight line running from top left to bottom right shows the thermal Jeans length as a function of column density 
and represents the expected dependence of filament width on density for self-gravitating filaments {\it in the absence of accretion}.
}
\label{coldens_width_acc}
\end{figure}

This accretion process likely plays a key role in effectively preventing dense filaments from collapsing to spindles
and allowing them to maintain approximately constant inner widths while evolving. 
Accretion supplies gravitational energy to supercritical filaments which is then converted into turbulent kinetic energy 
\citep[cf.][]{Heitsch+2009,KlessenHennebelle2010} 
and may explain the increase in velocity dispersion ($\sigma_{\rm tot}$) with 
column density observed by Arzoumanian et al. (2013). 
The central diameter of such accreting filaments is expected to be on the order the effective Jeans length 
$D_{\rm J,eff} \sim 2\, \sigma_{\rm tot}^2/G\Sigma_0 $, which Arzoumanian et al. (2013) have shown to remain close to $\sim 0.1$~pc. 
Toy models of this process are discussed by \citet{Heitsch2013} 
and \citet{HennebelleAndre2013}.
Given the importance of magnetic fields in the formation and evolution of filaments (see \S~6 below), 
the ``turbulence'' driven in the interior of dense filaments by the accretion of background matter is likely to be 
MHD in nature rather than purely hydrodynamic. It is therefore expected to dissipate due to ion-neutral friction. 
As shown  by Hennebelle \& Andr\'e (2013), a dynamical equilibrium can then be established 
between the supply and the dissipation of kinetic energy inside the filament, resulting in an almost constant inner width $\sim 0.1\,$pc 
in good agreement with observations 
(see Fig.~\ref{coldens_width_acc}).

\begin{figure}[!h]
\center
\resizebox{7.5cm}{!}{     
\includegraphics[angle=0,scale=0.5]{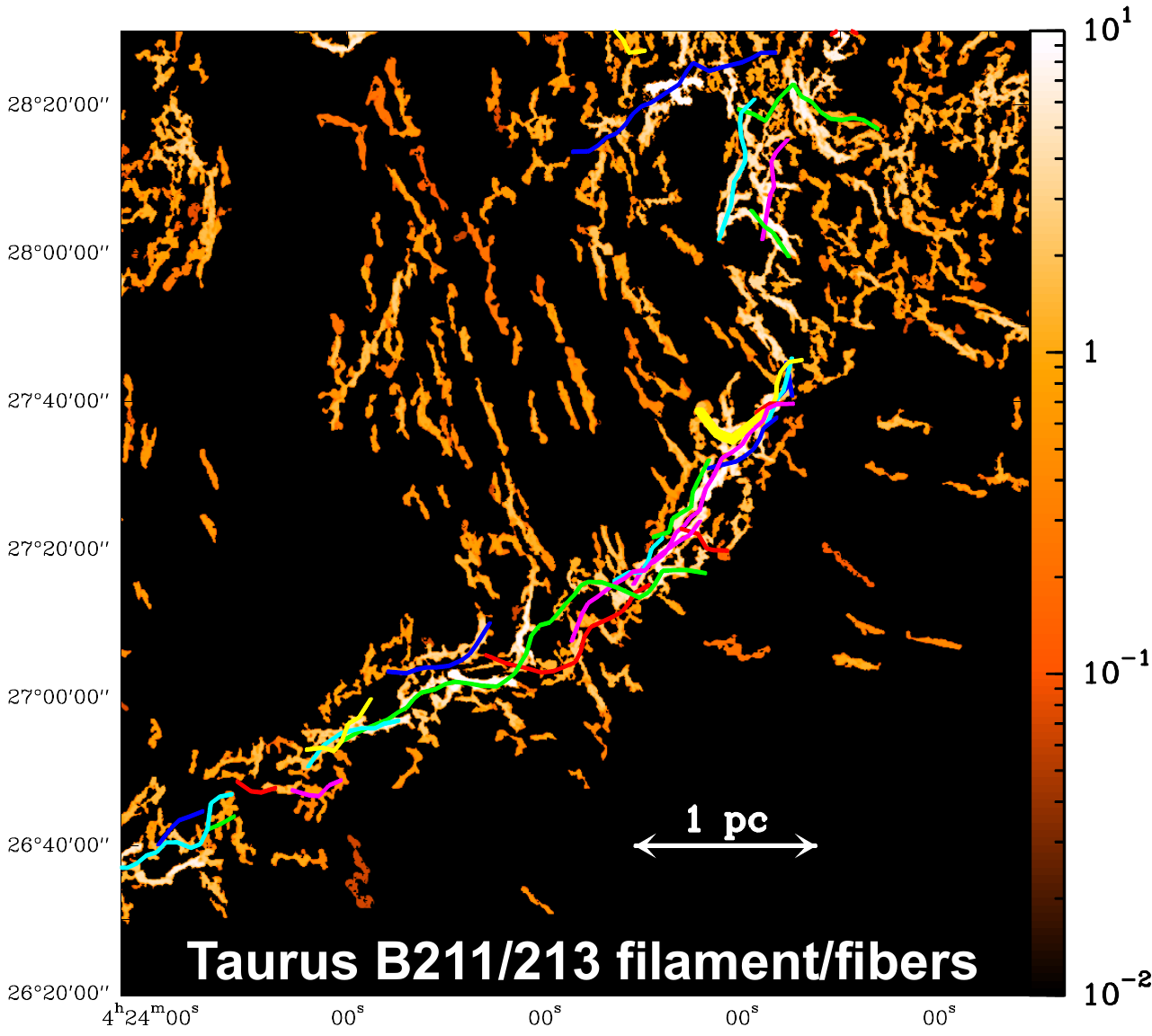}}
\caption{Fine (column) density structure of the B211/B213 filament based on a filtered version of the {\it Herschel} 250 $\mu$m 
    image of Palmeirim et al. (2013) using the 
    algorithm {\it getfilaments} \citep{Menshchikov2013}. In this view, all transverse angular scales larger than 72\arcsec ~(or $\sim 0.05$~pc) 
    were filtered out to enhance the contrast of the small-scale structure. 
    The color scale on the right is in MJy/sr at 250 $\mu$m. 
    The colored curves display the velocity-coherent fibers  
     identified by \citet{Hacar+2013} using C$^{18}$O and N$_2$H$^+$ line observations. }
\label{b211_fil_fibers}
\end{figure}

A possible manifestation of accretion-driven MHD turbulence may have been detected in the form of substructure and velocity-coherent 
``fibers'' in several star-forming filaments such as Taurus B211/B213 
\citep[e.g.][-- see Fig.~\ref{b211_fil_fibers}]{Hacar+2013,Tafalla+2015, 
Henshaw+2016}. 
Not all filaments consist of multiple fiber-like structures, however. The Musca filament, for instance, appears to be a 6-pc-long velocity-coherent 
sonic filament with much less substructure than the Taurus B211/B213 filament 
and no evidence for multiple  velocity-coherent fibers \citep{Hacar+2016,Cox+2016}. 
In the context of the accretion-driven turbulence picture introduced earlier, Cox et al. (2016) proposed that the Musca cloud may represent 
an earlier evolutionary stage that the Taurus B211/B213 filament in which the main filament has not yet accreted sufficient mass 
and energy to develop a multiple system of filamentary sub-components. A thermally supercritical filament may indeed be expected to 
develop a more complex system of intertwined fibers as the filament system grows in mass per unit length and its internal velocity dispersion 
increases.

\section{Role of magnetic fields}

\begin{figure}[!h]
\center
\resizebox{7.5cm}{!}{     
\includegraphics[angle=0,scale=0.5]{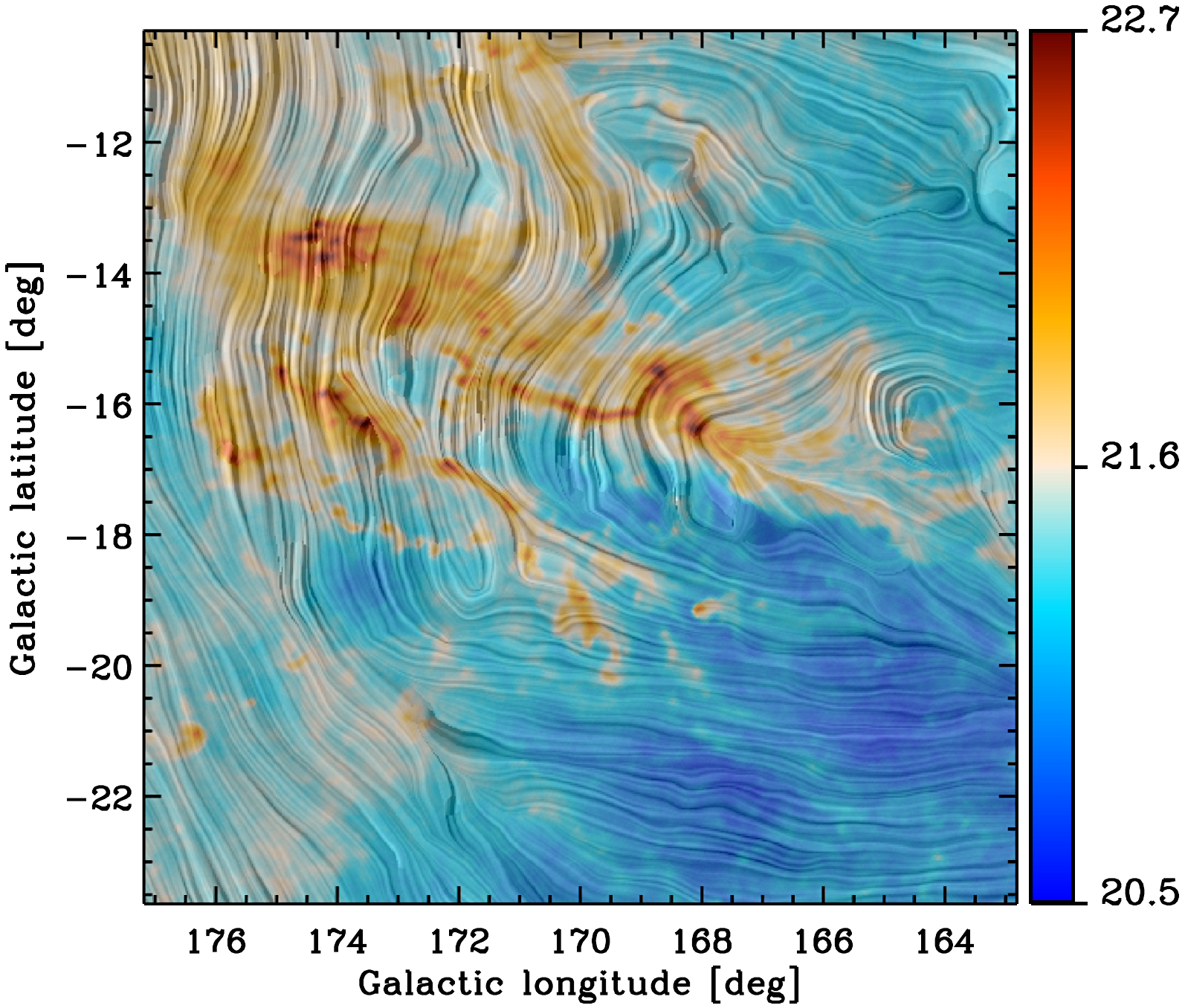}}
\caption{Magnetic field and column density map of the Taurus molecular cloud as measured by {\it Planck} 
(adapted from Planck Collaboration Int. XXXV, 2016). The color scale represents column density.
The superimposed  Ôdrapery patternÕ traces the magnetic-field orientation projected 
on the plane of the sky, as inferred from {\it Planck} polarization data at 850 $\mu $m. }
\label{taurus_planck}
\end{figure}

All-sky dust polarization observations with the {\it Planck} satellite 
at $850\, \mu$m have revealed very organized magnetic fields on large scales ($>\,$1--10~pc) in 
interstellar clouds \citep[][-- see Fig.~\ref{taurus_planck}]{PlanckXIX2015}. 
Low-density filamentary structures tend to be aligned with the magnetic field, 
while dense star-forming filaments tend to be perpendicular to the magnetic field \citep{PlanckXXXII2016,PlanckXXXV2016}, 
a trend also seen in optical and near-IR polarization observations \citep[][]{Chapman+2011,Palmeirim+2013,Panopoulou+2016,Soler+2016}. 
This trend has been quantified by the Planck Collaboration using the Histogram of Relative Orientations (HRO) and 
the unprecedented statistics provided by {\it Planck} dust polarization observations in nearby molecular clouds 
(Planck Collaboration Int. XXXV 2016). 
The HRO \citep{Soler+2013} is a statistical tool that measures the relative orientation 
between the plane-of-sky magnetic field $B_\bot $ and 
the local column density gradient. 
Using this technique, Soler and collaborators were able to show 
that a switch occurs at ${\rm log}_{10}(N_H/{\rm cm}^{-2}) = 21.9 \pm 0.2 $ between filamentary structures preferentially 
parallel to the magnetic field (low $N_H$) 
and filamentary structures preferentially 
perpendicular to the magnetic field  
(high $N_H$). 
Expressed in terms of $H_2$ column densities, and taking into account the fact that {\it Planck} underestimates the central 
column density of narrow $\sim 0.1\,$pc filaments due to beam dilution, 
the switch occurs at $N_{H_2} \approx 7 \times 10^{21} {\rm cm}^{-2}$, which 
corresponds to the critical mass per unit length for 10~K gas filaments of $\sim 0.1$~pc width.

There is also a hint from $Planck$ polarization observations of the nearest 
clouds 
that the direction of the magnetic field may change {\it within} dense filaments
from nearly perpendicular in the ambient cloud to more parallel  
in the filament interior \citep[][]{PlanckXXXIII2016}. 

All of these findings suggest that magnetic fields play a crucial role in the formation and evolution of star-forming filaments. 
In particular, they support the view that dense filaments form by accumulation of interstellar matter along field lines. 
Comparison with MHD simulations of molecular cloud evolution also shows that the {\it Planck} results are only consistent 
with models in which the turbulence is sub-Alfv\'enic or at most Alfv\'enic (Planck Collaboration Int. XXXV, 2016). 
Note that a similar conclusion had been reached by \citet{Li+2009} based on a comparison of the magnetic field orientation
inferred on large ($\sim \,$10--100$\,$pc) scales by optical polarimetry with the magnetic field orientation
inferred on small ($<< 1\, $pc) scales by ground-based submm polarimetry.

The low resolution of {\it Planck} polarization data (5\arcmin--10\arcmin$\,$at best or 0.2--0.4 pc in nearby clouds) is however insufficient 
to probe the organization of field lines in the $\sim 0.1\,$pc interior of filaments.  
Consequently, 
the detailed role of magnetic fields in the evolution of star-forming filaments and their fragmentation into prestellar cores 
remains a key open issue for the future.

\section{Conclusions and prospects}

The observational results discussed in this paper are quite encouraging as they point to a unified picture for star formation in 
in the molecular clouds of galaxies. 
They suggest that both the IMF (see \S~3) and the star formation efficiency in dense molecular gas 
\citep{Andre+2014,Shimajiri+2017} 
are largely set by the ``microphysics'' of core formation along filaments. 
Many open issues remain, however, especially concerning the detailed role of magnetic fields (see \S~6). 
High-resolution, high-dynamic range dust polarization observations at far-IR/submm wavelengths 
with future space observatories (possibly SPICA) 
should be able to test the hypothesis, tentatively suggested by {\it Planck} polarization results, 
that the B field becomes nearly parallel to the long axis of star-forming filaments in their dense interiors
on scales $< 0.1$~pc, due to, e.g., gravitational compression. 
Should this prove to be the case, it could explain both how dense filaments maintain a roughly constant $\sim 0.1\,$pc width while 
evolving 
and why the observed core spacing along filaments 
is significantly shorter than the 
characteristic fragmentation scale of 4$\, \times$ the filament diameter expected in the case of non-magnetized nearly isothermal gas cylinders 
(see \S~2.2). 
This would have profound implications for our understanding of core and star formation in the Universe. 

\section*{Acknowledgments}
I am grateful to my colleagues D. Arzoumanian, V. K\"onyves, A. Men'shchikov, 
P. Palmeirim, Y. Shimajiri, A. Bracco for their key contributions and   
to S. Inutsuka and P. Hennebelle 
for insightful discussions.  
This work was supported by the European Research Council 
under the European Union's Seventh Framework Programme 
(ERC Grant Agreement no. 291294 --  `ORISTARS').

\section*{References}


\bibliographystyle{aa}







\end{document}